\documentclass[wssquare]{ws-procs961x669}

\pdfoutput=1
\usepackage{mathtools,mathrsfs}
\usepackage{tensor}

\makeatletter
\newcommand*{\defeq}{\mathrel{\rlap{%
	\raisebox{0.3ex}{$\m@th\cdot$}}%
	\raisebox{-0.3ex}{$\m@th\cdot$}}%
	=}
\newcommand*{\eqdef}{=\mathrel{\rlap{%
	\raisebox{0.3ex}{$\m@th\cdot$}}%
	\raisebox{-0.3ex}{$\m@th\cdot$}}%
	}
\makeatother

\def\sg{\textsl{g}}

\usepackage{euscript}
\def\eE{\EuScript{E}}
\def\eF{\EuScript{F}}
\def\eL{\EuScript{L}}

\newcommand{\sumj}{\sum\limits_{j \geqslant \frac{1}{2}}^\infty}
\newcommand{\overbar}[1]{\mkern 1.5mu\overline{\mkern-1.5mu#1\mkern-1.5mu}\mkern 1.5mu}

\usepackage{xcolor}

\usepackage{hyperref}
\hypersetup{colorlinks,linkcolor={blue!55!black},citecolor={red!45!black},urlcolor={blue!45!black},breaklinks=true}

\begin{document}

\title{Constraining modified gravity theories with physical black holes}

\author{Sebastian Murk}
\address{Department of Physics and Astronomy, Macquarie University,\\
Sydney, New South Wales 2109, Australia\\
and \\
Sydney Quantum Academy,\\
Sydney, New South Wales 2006, Australia\\
E-mail: \href{mailto:sebastian.murk@mq.edu.au}{sebastian.murk@mq.edu.au}}

\begin{abstract}	
We review the constraints modified theories of gravity must satisfy to be compatible with the spherically symmetric black hole solutions of semiclassical gravity that describe the formation of an apparent horizon in finite time of a distant observer. The constraints are satisfied in generic modified gravity theories with up to fourth-order derivatives in the metric, indicating that the semiclassical solutions correspond to zeroth-order terms in perturbative solutions of these models. From an observational point of view, this result implies that it may not be possible to distinguish between the semiclassical theory and modifications including up to fourth-order derivatives based on the observation of an apparent horizon alone.
\end{abstract}

\keywords{black holes; general relativity; semiclassical gravity; modified gravity; alternative theories of gravity; quantum aspects of black holes.\\[3mm]
Contribution to the proceedings of the 16th Marcel Grossmann meeting (5--10 July 2021).\\
Session classification: Black holes in alternative theories of gravity}

\bodymatter
\thispagestyle{empty}

\section{Introduction} \label{sec:intro}
The predictions of general relativity (GR) have been confirmed in numerous experiments and are so far compatible with all currently available astrophysical and cosmological data. In particular, strong evidence for the existence of astrophysical black holes (ABHs) --- massive dark compact objects --- has accumulated over the last few decades. The precise nature of ABHs is still under debate, but contemporary models describe them as ultra-compact objects (UCOs) with or without a horizon \cite{bcns:19}. Due to their compactness, they provide excellent opportunities to probe strong gravity \cite{cp:19}.
 
The development of alternative theories of gravity has received much attention in recent years in an attempt to alleviate some of the perceived shortcomings of GR (such as the presence of singularities) and incorporate quantum gravitational effects \cite{b:04,dh:15}. As a preliminary test, any modified theory of gravity (MTG) should be compatible with the observed ABH candidates and provide a model to describe them. Present-day astronomical observations such as the detection of gravitational waves from coalescing compact systems \cite{ligo.virgo:16,ligo.virgo:19} and measurements of the M87 black hole shadow \cite{ehtc:20} probe the (dynamic) strong-field regime of gravity and limit the extent to which MTG can deviate from GR while still being compatible with observational data.

There is no single unanimously agreed upon definition of what exactly constitutes a black hole as different lines of research focus on different aspects. Consequently, the most useful and/or practical definition of a black hole depends on the context and highlights the relevant features. Nevertheless, a widely accepted feature of black holes is the presence of a trapped spacetime region from which nothing can escape \cite{c:19}. We adapt the terminology of Ref.~\citenum{f:14} and refer to a trapped region that is bounded by an apparent horizon as a physical black hole (PBH). The apparent horizon (see \sref{sec:preliminaries}) is an observer-dependent notion of the boundary of a black hole that is --- at least in principle --- physically observable (in contrast to the global, observer-independent notion of the event horizon) \cite{v:14}. It is therefore possible that GR and various MTG can be distinguished observationally based on the potentially distinct properties of PBHs and their horizons. 

\section{Mathematical preliminaries} \label{sec:preliminaries}
We work in units where $\hbar = c = G =1$, use the $(-+++)$ signature of the metric, and describe dynamics using the semiclassical Einstein equations $\tensor{G}{_\mu_\nu} = 8 \pi \tensor{T}{_\mu_\nu}$ or modifications thereof, where $\tensor{T}{_\mu_\nu} \equiv \langle \tensor{\hat{T}}{_\mu_\nu} \rangle_\omega$ on the rhs denotes the expectation value of the renormalized energy-momentum tensor (EMT) that describes both the collapsing matter and the produced excitations. No assumptions are made about the matter content of the theory, the quantum state $\omega$, and the status of various energy conditions.

The field equations of a generally covariant metric theory of gravity can be derived from its action \cite{p:11}. Here, we omit discussions of the matter Lagrangian and boundary conditions and focus on the gravitational action
\begin{align}
	\mathcal{S} = \int \sqrt{-\sg} \; \eL_\text{g} \; d^4 x 
	\label{eq:action}
\end{align}
determined by the gravitational Lagrangian density $\eL_\text{g}(\tensor{\sg}{^\mu^\nu},\tensor{R}{_\mu_\nu_\rho_\sigma})$. In classical GR and semiclassical gravity, it is strictly linear in the Ricci scalar, i.e.\ $\eL_\text{g}(\tensor{\sg}{^\mu^\nu},\tensor{R}{_\mu_\nu_\rho_\sigma}) = R$. Modifications typically include higher-order curvature corrections, which we organize according to powers of derivatives in the metric, i.e.\
\begin{align}
	\begin{aligned}
			\eL_\text{g} \sqrt{-\sg} &= \frac{{M_\text{Pl}}^2}{16\pi} \big( R + \lambda \eF(\tensor{\sg}{^\mu^\nu}, \tensor{R}{_\mu_\nu_\rho_\sigma}) \big) \\
			&= \frac{{M_\text{Pl}}^2}{16\pi} R + a_1 R^2 + a_2 \tensor{R}{_\mu_\nu} \tensor{R}{^\mu^\nu} + a_3 \tensor{R}{_\mu_\nu_\rho_\sigma} \tensor{R}{^\mu^\nu^\rho^\sigma} + \ldots ,
		\end{aligned}
		\label{eq:gravLagr}
\end{align}
where $\lambda$ sets the scale of the perturbative analysis, the cosmological constant term was omitted, $M_\text{Pl}$ denotes the Planck mass, and the coefficients $a_1$, $a_2$, $a_3$ are dimensionless. 

We limit our exposition to spherical symmetry and work with a general spherically symmetric metric in Schwarzschild coordinates given by
\begin{align}
	ds^2 = - e^{2h(t,r)} f(t,r) dt^2 + f(t,r)^{-1} dr^2 + r^2 d\Omega_2 ,
	\label{eq:metric}
\end{align}
where 
\begin{align}
	f(t,r) \defeq \partial_\mu r \partial^\mu r \equiv 1 - C(t,r)/r
\end{align}
provides a coordinate-independent definition of the Misner--Sharp mass $C(t,r)/2$ \cite{ms:64}, $r$ denotes the areal radius, and the function $h(t,r)$ plays the role of an integrating factor in coordinate transformations, such as
\begin{align}
	dt = e^{-h} \left( e^{h_+} dv - f^{-1} dr \right) .
\end{align}
Explicit calculations can be simplified by introducing the effective EMT components
\begin{align}
	\tensor{\tau}{_t} \defeq e^{-2h} \tensor{T}{_t_t} , \qquad \tensor{\tau}{_t^r} \defeq e^{-h} \tensor{T}{_t^r} , \qquad \tensor{\tau}{^r} \defeq \tensor{T}{^r^r} .
	\label{eq:effectiveEMTcomp}
\end{align}
The apparent horizon plays a central role in our analysis. Its outer component is located at the Schwarzschild radius $r_\sg(t)$ that is defined as the largest root of $f(t,r)=0$ \cite{fefhm:17}. Albeit an observer-dependent notion in general, it is unambiguously defined in spherical symmetry. To be of physical relevance, it must form in finite time according to the clock of a distant observer \cite{mt:21c,dmt:21}. Moreover, to maintain predictability of the theory, it must be a regular surface \cite{hv:book,pt:book}, i.e.\ the curvature scalars
\begin{align}
	\mathrm{T} \defeq \tensor{T}{^\mu_\mu} = - R / 8\pi + \mathcal{O}(\lambda) , \qquad 
	\mathfrak{T} \defeq \tensor{T}{^\mu^\nu}\tensor{T}{_\mu_\nu} = \tensor{R}{^\mu^\nu}\tensor{R}{_\mu_\nu} / 64 \pi^2 + \mathcal{O}(\lambda^2) ,
	\label{eq:curvature-scalars}
\end{align}
must be finite at $r=r_\sg$. Mathematically, this requirement is expressed as
\begin{align}
	\mathrm{T} &= \left( \tensor{\tau}{^r} - \tensor{\tau}{_t} \right) / f \; \to \; g_1(t) f^{k_1} , 
	\label{eq:reg1} \\
	\mathfrak{T} &= \left[ \left( \tensor{\tau}{_t} \right)^2 - 2 \left( \tensor{\tau}{_t^r} \right)^2 + \left( \tensor{\tau}{^r} \right)^2 \right] / f^2 \; \to \; g_2(t) f^{k_2} ,
	\label{eq:reg2}
\end{align}
for some functions $g_{1,2}(t)$ and $k_{1,2} \geqslant 0$. While there a infinitely many solutions in principle, only two distinct classes of dynamic solutions that satisfy the regularity requirement are self-consistent \cite{t:20,mt:21a}. With respect to the scaling behavior $\tensor{\tau}{} \sim f^k$ of the effective EMT components of \eref{eq:effectiveEMTcomp}, they correspond to the values $k \in \lbrace 0,1 \rbrace$, where $k \equiv k_1 \equiv k_2$. Both of them violate the null energy condition (NEC) in the vicinity of the outer apparent horizon \cite{bmmt:19,t:20,dmt:21}. We briefly summarize their properties in what follows.

\section{Semiclassical PBH solutions in spherical symmetry} \label{sec:semiclassicalPBHs}
In spherical symmetry, the relevant components $\tensor{G}{_t_t}$, $\tensor{G}{_t^r}$, $\tensor{G}{^r^r}$ of the semiclassical Einstein equations can be written as
\begin{align}
	\partial_r C &= 8 \pi r^2 \tensor{\tau}{_t} / f ,
	\label{eq:EEGRtt} \\
	\partial_t C &= 8 \pi r^2 e^{h} \tensor{\tau}{_t^r} ,
	\label{eq:EEGRtr} \\
	\partial_r h &= 4 \pi r \left( \tensor{\tau}{_t} + \tensor{\tau}{^r} \right) / f^2 ,
	\label{eq:EEGRrr}
\end{align}
respectively. 

\subsection{$k=0$ solutions} \label{subsec:k0PBHs}
Close to the horizon (i.e.\ as $r \to r_\sg$), the limiting form of the effective EMT components is given by
\begin{align}
	\tensor{\tau}{_t} = - \Upsilon^2 + \sumj e_j x^j , \qquad 
	\tensor{\tau}{_t^r} = \pm \Upsilon^2 + \sumj \phi_j x^j , \qquad
	\tensor{\tau}{^r} = - \Upsilon^2 + \sumj p_j x^j ,
\end{align}
for some time-dependent function $\Upsilon(t) > 0$, where $j \in \mathbb{Z} \frac{1}{2}$ labels half-integer and integer coefficients and powers of the coordinate distance $x \defeq r - r_\sg$ from the apparent horizon, and the lower (upper) signature of $\tensor{\tau}{_t^r}$ corresponds to a contracting (an expanding) Schwarzschild radius $r_\sg$. Taking $\tensor{\tau}{_t} \to \tensor{\tau}{^r} \to + \Upsilon^2 + \mathcal{O}(\sqrt{x})$ results in complex-valued solutions of the Einstein equations \eqref{eq:EEGRtt}--\eqref{eq:EEGRrr} \cite{bmmt:19}. The leading terms of the metric functions that solve Eqs.~\eqref{eq:EEGRtt} and \eqref{eq:EEGRrr} are given by
\begin{align}
	C &= r_\sg - c_{12} \sqrt{x} + \sum\limits_{j \geqslant 1}^\infty c_j x^j , \\
	h &= - \frac{1}{2} \ln \frac{x}{\xi} + \sumj h_j x^j ,
\end{align}
where the leading coefficient $c_{12} = 4 \sqrt{\pi} r_\sg^{3/2} \Upsilon$, the function $\xi(t)$ is determined by the choice of time variable, and higher-order terms must be matched with those of the EMT expansion. Eq.~\eqref{eq:EEGRtr} then necessitates that the dynamic behavior of the horizon is governed by 
\begin{align}
	r_\sg^\prime &= \pm c_{12} \sqrt{\xi} / r_\sg ,
\end{align}
where the lower (upper) signature describes an evaporating PBH (an expanding white hole).

\subsection{$k=1$ solution} \label{subsec:k1PBHs}
For $k=1$, the limiting form of the effective EMT components close to the horizon (i.e.\ as $r \to r_\sg$) is given by
\begin{align}
	\tensor{\tau}{_t} = E f + \sum\limits_{j \geqslant 2}^\infty e_j x^j , \qquad \tensor{\tau}{_t^r} = \Phi f + \sum\limits_{j \geqslant 2}^\infty \phi_j x^j , \qquad \tensor{\tau}{^r} = P f + \sum\limits_{j \geqslant 2}^\infty p_j x^j ,
	\label{eq:k1taus}
\end{align}
where $E$, $\Phi$, and $P$ denote the energy density, flux, and pressure at the horizon. The leading terms of the metric functions that solve Eqs.~\eqref{eq:EEGRtt} and \eqref{eq:EEGRrr} are given by
\begin{align}
	C &= r - c_{32} x^{3/2} + \mathcal{O}(x^2) , \\
	h &= - \frac{3}{2} \ln \frac{x}{\xi} + \mathcal{O}(\sqrt{x}) ,
\end{align}
where the leading coefficient $c_{32} = 4 r_\sg^{3/2} \sqrt{- \pi e_2 / 3}$, and $e_2$ denotes the $\mathcal{O}(x^2)$ coefficient of $\tensor{\tau}{_t}$, see Eq.~\eqref{eq:k1taus}. According to Eq.~\eqref{eq:EEGRtr}, the dynamic behavior of the horizon is governed by 
\begin{align}
	r_\sg^\prime &= \pm c_{32} \xi^{3/2} / r_\sg ,
\end{align}
where once again the lower (upper) signature describes an evaporating PBH (an expanding white hole). Since we are interested in scenarios resulting from gravitational collapse, we focus on evaporating PBH solutions ($r_\sg^\prime < 0$) in what follows. 

For $k=1$, only one solution is self-consistent \cite{mt:21a}. It describes PBHs at the instant of their formation and is uniquely defined by the energy density $E \defeq - \tensor{T}{^t_t}$ and pressure $P \defeq \tensor{T}{^r_r}$ at the horizon, which take on their maximal possible values $E = - P = 1/(8 \pi r_\sg^2)$. Hence, we refer to it as extreme-valued $k=1$ solution.

\section{Modified Einstein equations in spherical symmetry} \label{sec:MTG}
The modified Einstein equations are obtained through variation of the gravitational action \cite{p:11} [cf.\ Eqs.~\eqref{eq:action}--\eqref{eq:gravLagr}], and can be represented schematically as
\begin{align}
	\tensor{G}{_\mu_\nu} + \lambda \tensor{\EuScript{E}}{_\mu_\nu} = 8 \pi \tensor{T}{_\mu_\nu} , \label{eq:mEE}
\end{align}
where the terms $\tensor{\EuScript{E}}{_\mu_\nu}$ denote deviations from the Einstein equations that result from the variation of $\eF(\tensor{\sg}{^\mu^\nu}, \tensor{R}{_\mu_\nu_\rho_\sigma})$ [cf.\ Eq.~\eqref{eq:gravLagr}]. The explicit form of the Einstein equations in MTG then immediately follows from Eqs.~\eqref{eq:EEGRtt}--\eqref{eq:EEGRrr}:
\begin{align}
	& f r^{-2} e^{2h} \partial_r C + \lambda \tensor{\EuScript{E}}{_t_t} = 8 \pi \tensor{T}{_t_t} ,
	\label{eq:mEEtt} \\
	& r^{-2} \partial_t C + \lambda \tensor{\EuScript{E}}{_t^r} = 8 \pi \tensor{T}{_t^r} ,
	\label{eq:mEEtr} \\
	& 2 f^2 r^{-1} \partial_r h - f r^{-2} \partial_r C + \lambda \tensor{\EuScript{E}}{^r^r} = 8 \pi \tensor{T}{^r^r} .
	\label{eq:mEErr}
\end{align}
The metric functions that solve the modified Einstein equations \eqref{eq:mEE} are
\begin{align}
	C_\lambda & \eqdef \bar{C}(t,r) + \lambda \Sigma(t,r) , 
	\label{eq:C_lambda} \\
	h_\lambda & \eqdef \bar{h}(t,r) + \lambda \Omega(t,r) ,
	\label{eq:h_lambda}
\end{align}
where $\Sigma$ and $\Omega$ denote the perturbative corrections, and the bar labels the semiclassical metric functions introduced in \sref{sec:semiclassicalPBHs}. Schematically, substitution of Eqs.~\eqref{eq:C_lambda}--\eqref{eq:h_lambda} into the modified Einstein equations \eqref{eq:mEE} yields equations of the form
\begin{align}
	\tensor{\bar{G}}{_\mu_\nu} + \lambda \tensor{\tilde{G}}{_\mu_\nu} + \lambda \tensor{\bar{\eE}}{_\mu_\nu} = 8 \pi \left( \tensor{\bar{T}}{_\mu_\nu} + \lambda \tensor{\tilde{T}}{_\mu_\nu} \right) ,
\end{align}
where objects labeled by the tilde correspond to first-order terms in the expansion of $\lambda$ that involve the perturbative corrections $\Sigma$ and/or $\Omega$, and terms of order $\mathcal{O}(\lambda^2)$ and higher have been omitted. The modified gravity terms can be expressed solely in terms of unperturbed quantities, i.e.\ $\tensor{\bar{\eE}}{_\mu_\nu} \equiv \tensor{\eE}{_\mu_\nu}[\bar{C},\bar{h}]$. For a detailed account of the perturbative analysis, the reader is referred to Refs.~\citenum{mt:21b,m:21}.

\section{Constraints for modified gravity theories}
Any arbitrary MTG must satisfy several constraints \cite{mt:21b} to be compatible with the spherically symmetric PBH solutions of semiclassical gravity described in \sref{sec:semiclassicalPBHs}. The constraints are twofold: first, the terms $\tensor{\bar{\eE}}{_\mu_\nu}$ that encode deviations from the semiclassical theory must follow a particular structure when expanded in terms of the coordinate distance $x \defeq r - r_\sg$ from the apparent horizon. Second, several identities between their coefficients must hold.

To be compatible with PBH solutions of the $k=0$ type (\sref{subsec:k0PBHs}), the MTG terms $\tensor{\bar{\eE}}{_\mu_\nu}$ must conform to the expansion structures prescribed by
\begin{align}
	\tensor{\bar{\eE}}{_t_t} &= \frac{\ae_{\bar{1}}}{x} + \frac{\ae_{\overbar{12}}}{\sqrt{x}} + \ae_0 + \sumj \ae_j x^j , 
	\label{eq:k0-Ett} \\
	\tensor{\bar{\eE}}{_t^r} &= \frac{\oe_{\overbar{12}}}{\sqrt{x}} + \oe_0 + \sumj \oe_j x^j , 
	\label{eq:k0-Etr} \\
	\tensor{\bar{\eE}}{^r^r} &= \o_0 + \sumj \o_j x^j , 
	 \label{eq:k0-Err}
\end{align}
and the three identities
\begin{align}
	\ae_{\bar{1}} &= \sqrt{\bar{\xi}} \oe_{\overbar{12}} = \bar{\xi} \o_0 , \qquad \ae_{\overbar{12}} = 2 \sqrt{\bar{\xi}} \oe_0 - \bar{\xi} \o_{12} ,
	\label{eq:k0-con}
\end{align}
between their lowest-order coefficients must be satisfied. Again, the bar signifies that the MTG terms $\tensor{\bar{\eE}}{_\mu_\nu} \equiv \tensor{\eE}{_\mu_\nu}[\bar{C},\bar{h}]$ (and also $\bar{\xi} \equiv \xi[\bar{C},\bar{h}]$) can be expressed solely in terms of unperturbed solutions.

Similarly, to be compatible with the extreme-valued $k=1$ solution (\sref{subsec:k1PBHs}), the MTG terms must follow the expansion structures prescribed by
\begin{align}
	\tensor{\bar{\eE}}{_t_t} &= \frac{\ae_{\overbar{32}}}{x^{3/2}} + \frac{\ae_{\bar{1}}}{x} + \frac{\ae_{\overbar{12}}}{\sqrt{x}}  + \ae_0 + \sumj \ae_j x^j , 
	\label{eq:k1-Ett} \\
	\tensor{\bar{\eE}}{_t^r} &= \oe_0 + \sumj \oe_j x^j , 
	\label{eq:k1-Etr} \\
	\tensor{\bar{\eE}}{^r^r} &= \sum\limits_{j \geqslant \frac{3}{2}}^\infty \o_j x^j ,
	\label{eq:k1-Err}
\end{align}
and the two identities
\begin{align}
	\ae_{\overbar{32}} &= 2 \bar{\xi}^{3/2} \oe_0 - \bar{\xi}^3 \o_{32} , \qquad 
	\ae_{\bar{1}} = 2 \bar{\xi}^{3/2} \left( h_{12} \oe_0 + \oe_{12} \right) - \bar{\xi}^3 \left( 2 h_{12} \o_{32} + \o_2 \right) ,
	\label{eq:k1-con}
\end{align}
between their lowest-order coefficients must hold.

For both types of solutions $k \in \lbrace 0,1 \rbrace$, no additional constraints can be obtained through the consideration of higher-order terms ($\geqslant \mathcal{O}(x^{3/2})$) in the metric functions as in this case the modified Einstein equations contain too many additional independent variables. A comprehensive analysis of this result and a detailed derivation of Eqs.~\eqref{eq:k0-Ett}--\eqref{eq:k1-con} is provided in Ref.~\citenum{mt:21b}.

If a particular MTG does not satisfy the constraints for one or both classes of semiclassical PBHs, the theory in question may still possess solutions corresponding to PBHs, but their underlying mathematical description must then differ from that of \sref{subsec:k0PBHs} and/or \sref{subsec:k1PBHs}, which may or may not lead to observable differences based on properties of the associated near-horizon geometry.

So far, the constraints imposed by Eqs.~\eqref{eq:k0-Ett}--\eqref{eq:k1-con} have been investigated in MTG with up to fourth-order derivatives in the metric \cite{m:21}. In particular, it has been demonstrated that
\begin{enumerate}
	\item generic $\mathfrak{f}(R)$ theories, where $\eL_\text{g} = \mathfrak{f}(R)$, $\mathfrak{f}(R) \eqdef R + \lambda \eF(R)$ [cf.\ Eq.~\eqref{eq:gravLagr}], $\eF(R) = \beta R^q$, and $\beta, q \in \mathbb{R}$
	\item generic MTG with up to fourth-order derivatives in the metric, where the particular choice of $\eL_\text{g} = (- \alpha \tensor{R}{_\mu_\nu} \tensor{R}{^\mu^\nu} + \beta R^2 + \gamma \kappa^{-2} R)$ in the action is motivated by the desire to avoid ghosts (massive states of negative norm that result in an apparent lack of unitarity) \cite{s:78,sf:10}, and $\kappa^2 = 32 \pi$
\end{enumerate}
satisfy all of the constrains identically (i.e.\ without requiring any additional conditions) and are thus compatible with the existence of spherically symmetric semiclassical PBHs. The procedure for obtaining explicit expressions for the MTG terms $\tensor{\bar{\eE}}{_\mu_\nu}$ in terms of unperturbed quantities is outlined in Ref.~\citenum{m:21}, and explicit expressions for the relevant MTG coefficients of Eqs.~\eqref{eq:k0-con} and \eqref{eq:k1-con} in these two theories are provided in the following {\href{https://github.com/s-murk/MTGcoefficients}{linked Github repository}}.

\section{Discussion}
We have outlined a procedure to test the compatibility of arbitrary metric MTG with the PBH solutions of semiclassical gravity, which are characterized by the formation of a regular apparent horizon in finite time of a distant observer. We find that generic MTG with up to fourth-order derivatives in the metric identically satisfy all of the constraints for both classes of solutions, indicating that the semiclassical PBHs correspond to zeroth-order terms in perturbative solutions of these models. As a result, the observation of an apparent horizon by itself may not be sufficient to distinguish between the semiclassical theory and modifications including up to fourth-order derivatives in the metric. A detailed analysis of the response of the near-horizon geometry to perturbations is required, and may allow to identify potentially observable differences between the PBH solutions of various theoretical frameworks.

\section*{Acknowledgments}
I would like to thank Daniel Terno for useful discussions and helpful comments. This work was supported by an International Macquarie University Research Excellence Scholarship and a Sydney Quantum Academy Scholarship.

\end{document}